\documentclass[
aps,
prl,
twocolumn,
superscriptaddress
]{revtex4-2}

\usepackage[a4paper, total={7in, 9in}]{geometry}

\usepackage{graphicx}
\usepackage{amsmath}
\usepackage{mathrsfs}
\usepackage{setspace}
\usepackage{color}
\usepackage[dvipsnames,svgname]{xcolor}
\usepackage{textcomp}
\usepackage{booktabs}
\usepackage[flushleft]{threeparttable}
\usepackage{comment}
\usepackage{amssymb}
\usepackage{amsmath}
\usepackage{float}
\usepackage{psfrag}
\usepackage{siunitx}
\usepackage{romannum}
\usepackage{bm}

\def\be{\begin{equation}}
\def\ee{\end{equation}}
\def\ba{\begin{eqnarray}}
\def\ea{\end{eqnarray}}

\usepackage{esdiff}
\newcommand{\bsb}[1]{\boldsymbol{#1}}
\newcommand{\me}{\mathrm{e}}
\newcommand{\mi}{\mathrm{i}}

\begin{document}
\pagenumbering{arabic} 
\title{Jet Size Prediction in Compound Multiphase Bubble Bursting}
%bubble bursting with surface viscosity?

\author{Zhengyu Yang}
\thanks{These authors contributed equally.}
\affiliation{Mechanical Science and Engineering, University of Illinois at Urbana-Champaign, Illinois 61801, USA}

\author{Yang Liu}
\thanks{These authors contributed equally.}
\affiliation{Department of Engineering Mechanics, Tsinghua University, Beijing 100084, China}

\author{Jie Feng}
\email{jiefeng@illinois.edu}
\affiliation{Mechanical Science and Engineering, University of Illinois at Urbana-Champaign, Illinois 61801, USA}
\affiliation{Materials Research Laboratory, University of Illinois at Urbana-Champaign, Illinois 61801, USA}
\begin{abstract} 
The Worthington jets from bursting bubbles at a gas-liquid interface can break up into small droplets, aerosolizing chemical and biological substances into the atmosphere and impacting both global climate and public health. Despite their importance in contaminant transport, the influence of adsorbed contaminants on bubble-bursting jet dynamics remains poorly understood. Here, we document how an immiscible compound contaminant layer impacts the jet radius, which deviates from the expected jetting dynamics produced by clean bubble bursting. We rationalize the deviation of the jet radius by characterizing the propagation of the capillary waves at the air-oil-water interface. We develop a linearized wave damping model based on the understanding of the oil thickness profile and the wave dispersion, and we propose a revised Ohnesorge number with a scaling relation that captures the experimental results reasonably well across a wide range of oil layer thicknesses and viscosities. Our work not only advances the fundamental understanding of bubble bursting jets but also offers valuable insights for predicting aerosol size distributions and modeling the transport of airborne contaminants in realistic environmental scenarios.
\end{abstract}

\maketitle
\newpage

\section*{Introduction}
Bubbles bursting at an air-liquid interface are ubiquitous across a broad range of natural and industrial processes \citep{bird2010daughter,feng2014,veron2015ocean,deike2022mass,bourouiba2021fluid}. The bursting ejects a multitude of drops and plays a pivotal role in mass transport across the air–liquid interface, as the ejection of tiny drops strongly influences phenomena such as sea spray aerosol production \cite{deike2022mass}, air pollution \citep{murphy2016depth,trainic2020airborne}, global climate \citep{wang2017role, veron2015ocean, wilson2015}, and even transmission of infectious diseases \cite{bourouiba2020}. In this context, drops produced by the pinching-off of Worthington jets from bursting bubbles have received significant research attention. 
Most previous studies have focused on bursting jets of clean bubbles \cite{deike2018dynamics,brasz2018minimum,ganan2021physics,gordillo2023theory}, yet bubbles with compound interfaces are far more common in practical scenarios. For example, rising bubbles in water bodies often scavenge dispersed substances that they collide with, resulting in contaminated bubble-liquid interfaces \cite{walls2014moving,ji2022water}. Such compound bubbles are prevalent in processes like natural gas seeps  \cite{johansen2017time}, oil spills \cite{ehrenhauser2014bubble}, froth flotation \citep{zhou2014application,su2006role}, and material processing \citep{behrens2020oil,visser2019architected}. Despite their widespread occurrence, the influence of these contaminated compound interfaces on bubble bursting dynamics remains largely unexplored, and no scaling relation is available to predict the jet drop size from compound bubbles to the best of our knowledge.

The formation of Worthington jets during bubble bursting is driven by the radially inward interfacial capillary waves excited after the film rupture \cite{deike2018dynamics,brasz2018minimum,gordillo2019capillary}.
Consequently, contaminants adsorbed to the bubble surface are expected to impact the kinematics of the jets significantly by affecting interfacial capillary wave propagation. Rich and distinct dynamics have been observed in previous investigations for contaminated bubble bursting. For instance, the suppression of jets has been shown in bursting bubbles with interfaces laden by surfactants \cite{pierre2022influence,vega2024influence} or proteins \cite{ji2023secondary,yang2024effect}, due to the capillary wave propagation being impeded by Marangoni or surface viscoelastic effects, respectively. In contrast, for a contaminated bubble coated by an immiscible compound layer, a thin and fast jet emerges as the compound interface enhances the damping of the capillary wave, ejecting jet drops with a much smaller size \cite{yang2023enhanced}. Given the ubiquity and peculiarity of bubble bursting dynamics with non-bare interfaces, a fundamental framework is required to rationalize the influence of contaminants on jet formation and size prediction.

Here, with a representative oil-coated bubble configuration, we aim to develop a scaling relation to predict the jet size by quantifying the influence of the immiscible compound layer.
We focus on the damping features of the capillary waves during their propagation at the surface of the collapsing bubble cavity. With experimental support, we employ a linearized capillary wave model to rationalize their dispersion and dissipation at the compound bubble surface. A scaling relation based on the revised Ohnesorge number for the jet radius is proposed to capture the experimental trend. 

\section*{Dynamics of oil-coated bubble bursting}
\begin{figure*} 
    \centering
    \includegraphics[width=\linewidth]{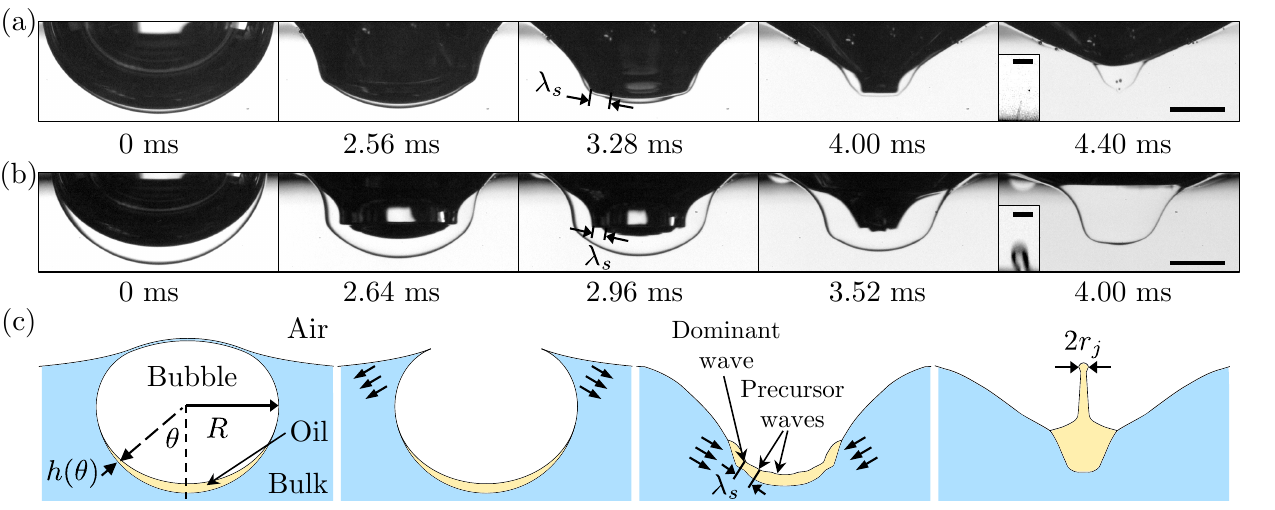}
    \caption{ (a, b) Cavity collapse during the bursting of bubbles coated by a 0.9 mPa s silicone oil layer in a glycerin solution of viscosity 9.5 mPa s. The corresponding oil fractions $\psi_o$ are (a) 0.6\% and (b) 8.2\%, respectively (See SI Videos 1 and 2). The capillary waves propagate down the cavity and finally focus at the cavity bottom, producing an upward jet. The wavelength $\lambda_s$ of the secondary wave, which is between the rearmost dominant wave trough and the closest wave trough, is marked and decreases as $\psi_0$ increases. The inset shows the upward jet when the jet tip reaches the undisturbed air-water level. 
     The scale bars represent 1 mm and 200 $\mu$m for the cavity in the main figure and the jet in the inset, respectively. 
    (c) Schematics of bubble bursting producing a jet. During the bursting of an oil-coated bubble, the bubble cap first ruptures and initiates capillary waves downward along the bubble surface. The train of capillary waves contains the rearmost and most energetic dominant wave specified as the dominant wave. We denote the capillary waves excited in front of the dominant wave with a shorter wavelength specified as precursor waves. The jet is produced as a result of dominant wave focusing, overcoming the perturbations of the precursor waves.}
    \label{fig:images}
\end{figure*}

Figure \ref{fig:images} (a,b) shows the high-speed images for the bursting dynamics of oil-coated bubbles with different oil volume fractions $\psi_o = 3V_o/(4\pi R^3)$ while keeping the same liquid combination. Here, $V_o$ represents the oil volume and $R$ represents the bubble radius.
The capillary waves are initiated after the bubble cap film ruptures. The waves travel down the cavity with their crests moving in phase on both the air–oil and oil–water interfaces, sweeping the oil toward the nadir of the cavity. As the capillary waves encounter a thicker oil layer during their propagation, they split onto oil-air and oil-water interfaces. The capillary waves reach the cavity bottom more quickly at the oil-air interface with a shorter traveling distance than those at the oil–water interface. The capillary waves at the oil-air interface converge and form a truncated cone-shaped cavity within the oil layer, ultimately producing an upward oily jet by the curvature reversal of the cavity bottom. Such dynamics of jet formation is summarized in Figure~\ref{fig:images} (c).

Characterizing the capillary wave propagation during bubble bursting is critical for describing the jet dynamics \cite{krishnan2017scaling,gordillo2019capillary,yang2023enhanced}. Within the train of capillary waves, we denote the last and most energetic capillary wave as the dominant wave. Meanwhile, the capillary waves preceding the dominant wave, which travel faster with shorter wavelengths, are referred to as precursor waves. Specifically, the eminent wave closest to the dominant wave is considered as an indicator for wave damping due to its longest wavelength among the precursor waves, and we denote this capillary wave as the secondary wave with wavelength $\lambda_s$.
For bubbles in a high-viscosity bulk liquid and coated by a thinner layer of low-viscosity oil (Fig. \ref{fig:images}(a)), we observe that the cavity bottom is relatively smooth and flat with weaker precursor waves before the dominant wave converges. It suggests that the damping of the precursor waves is stronger, and therefore, the cavity collapse approaches the singular limit for a smaller jet radius $r_j$ \cite{krishnan2017scaling,gordillo2019capillary}. Consistently, a thin jet is observed (inset of Fig. \ref{fig:images}(a)). In comparison, the precursor waves are more prominent and shorter in wavelength $\lambda_s$ with a thicker layer of low-viscosity oil as shown in Fig. \ref{fig:images}(b). The significant perturbations from the stronger precursor waves produce a larger $r_j$. In our experiments, we note that this change in wave damping and jet size in the case of $\mu_o/\mu_w<1$ noticeably differs from the cases of $\mu_o/\mu_w>1$, where an increase in $\psi_o$ enhances viscous damping and results in a smaller $r_j$. 
The above observations demonstrate that the compound interface fundamentally alters jetting dynamics by modulating capillary wave damping, which necessitates careful characterization and interpretation. 

\section*{Damping of capillary waves}

The jetting dynamics of bare bubble bursting at an air-liquid interface is determined by two dimensionless numbers: Ohnesorge number $Oh=\mu/\sqrt{\rho\gamma R}$ and Bond number $Bo = \rho gR^2/\gamma$, where $\gamma, \rho$ and $\mu$ are the surface tension, density, and viscosity of the liquid, respectively. The Bond number, which compares the gravity and capillary effects, is sufficiently small in our experiments, and therefore the gravity effect is negligible \cite{deike2018dynamics}. Therefore, the Ohnesorge number $Oh = t_c/t_d$ (comparing the inertio–capillary to inertio–viscous timescales) controls the jet dynamics.
Here, $t_c = \sqrt{(\rho R^3)/\gamma}$ represents the inertio-capillary timescale, and $t_d =\rho R^2/\mu$ represents the inertio–viscous timescale. Since the amplitude damping of capillary waves controls the ejection of bubble bursting jets \cite{krishnan2017scaling}, $Oh$ can also be interpreted as a description for the progressive damping of the precursor waves during the bubble cavity collapse. The viscous damping rate for the amplitude of a capillary wave with wavelength $\lambda_s$ at a free liquid surface can be estimated as $T_{\lambda}^{-1}{=2\nu k^2 = 8\pi^2\mu/(\rho \lambda_s^2)}$. Here, $\nu=\mu/\rho$ represents the kinematic viscosity, and $k = 2\pi/\lambda_s$ represents the wavenumber. Given that $\lambda_s \propto R$ in bubble bursting, $T_\lambda^{-1}\propto t_d^{-1}$. Additionally, the cavity collapse occurs on a timescale of $t_c$, as the time for the capillary waves to reach the bottom of the cavity $t_{bc} \propto t_c$. Consequently, the total viscous damping of the precursor waves is characterized by $T_\lambda^{-1}t_{bc} \propto Oh$. Essentially, a higher $Oh$ indicates stronger damping of the precursor waves during the cavity collapse, sheltering the energy focusing of the dominant wave from the perturbation of the precursor waves, and finally resulting in thinner and faster jets \cite{krishnan2017scaling,deike2018dynamics,brasz2018minimum,gordillo2019capillary}.

\begin{figure}
    \centering
    \includegraphics[width=\linewidth]{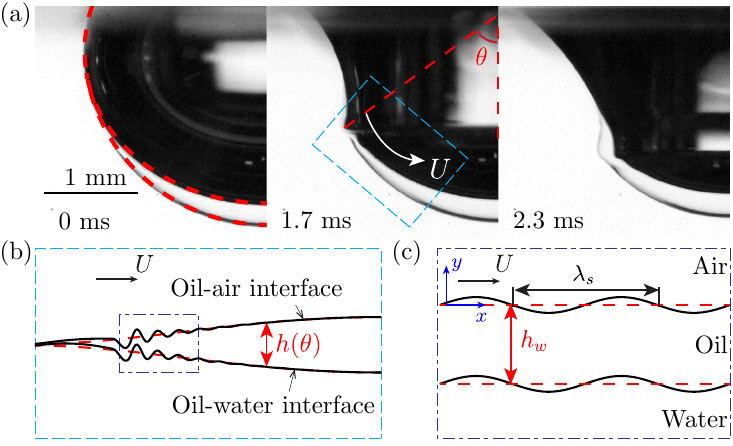}
    \caption{(a) Capillary waves propagating down an oil-coated bubble with parameters of $\mu_o$ = 4.6 mPa s, $\mu_w$ = 2.1 mPa s, and $\psi_o$ = 5.1\%. The red dashed curve indicates the shape of the oil-coated bubble calculated using Young-Laplace equations. (b) Schematics of the capillary waves propagating at an increasing oil layer thickness $h(\theta)$. The surface is flattened to illustrate the capillary waves propagating toward the thicker part of the oil layer. (c) Schematic of the 2D planar wave model at an oil-coated air-water interface for calculation of the viscous damping rate of the precursor waves. }
    \label{fig:cavity}
\end{figure}

For an oil-coated bubble, however, the damping of capillary waves is influenced by the multiphase compound interface (Fig. \ref{fig:cavity}(a)). 
Considering a static oil-coated bubble at the water surface, the oil layer is thicker near the bubble bottom due to the balance between surface tension and gravity. The capillary waves encounter an oil layer with an increasing thickness $h(\theta)$ as they propagate towards the cavity bottom with the polar angle $\theta$ decreasing, as shown in the schematic of Fig. \ref{fig:cavity}(b). Therefore, the above-mentioned capillary wave analysis for bare bubbles, which considers only the bulk liquid properties, requires careful revision for the compound bubbles. During capillary wave propagation, the energy dissipation arises from the viscous resistance to motions within both the oil layer and the bulk liquid. In particular, the damping of precursor capillary waves is time-dependent because of the non-uniform oil layer thickness. As the wave propagates into a thicker oil layer, the oil layer essentially contributes more to the total dissipation. This is in contrast to the constant damping rate estimated for the clean interface in the case of bare bubble bursting, which motivates us to furnish the capillary wave model to account for the difference between the compound and clean interfaces. 

We aim to propose a minimal capillary wave model to describe this time-dependent capillary wave damping considering the following conditions and assumptions. In our experiments, the typical wavelength $\lambda_s\approx 0.2-0.4 $ mm is an order of magnitude smaller than the capillary length $\lambda_c=\sqrt{\gamma_{oa}/(\rho_og)}$, so the contribution of gravity becomes negligible \citep{lamb1924hydrodynamics}. Additionally, the surface curvature effect of the bubble radius is negligible since $\lambda_s/R\ll 1$ \cite{lamb1924hydrodynamics}. Therefore, we apply a 2D planar wave model to characterize the capillary waves on the cavity surface. The non-linear effects from the viscosity are considered negligible as our wavelength is much larger than the interfacial boundary layer thickness $l_\nu=\sqrt{2\mu_o/(\rho_o \omega_0)} \approx 0.02$ mm \cite{denner2016frequency,shen2017marangoni}. Here, $\omega_0$ represents the oscillation frequency of the wave. Furthermore, the change in oil layer thickness within a wave period is insignificant considering $\lambda_s/R\ll 1$ and $h/R\ll 1$, allowing us to model a linear, periodic 2D capillary wave with a local oil layer thickness $h_w$ experienced by the precursor waves as shown in Fig. 
\ref{fig:cavity}(c).

Following the above assumptions, the capillary wave motion in both phases satisfies the linearized Navier-Stokes equation for Newtonian liquids as
\begin{equation} \label{eq:ns}
    \rho_{w,o}\diffp{\bsb{u}_{w,o}}{t}=-\nabla p_{w,o} + \mu_{w,o}\nabla^2\bsb{u}_{w,o},
\end{equation}
Here, $p_{w,o}$ represents pressure, and $\bsb{u}_{w,o}$ is a 2D periodic velocity field of
\begin{equation} \label{eq:normalmode}
    \bsb{u}_{w,o} = \bsb{U}_{w,o}(y)\me^{\mi(2\pi x/\lambda_s+\omega t)}.
\end{equation} 
The subscript $w,o$ indicates that Eqs. (\ref{eq:ns},\ref{eq:normalmode}) apply to the motion within the domain of both water ($y<-h_w$) and oil ($-h_w<y<0$).
$\bsb{U}(y)$ represents the velocity dependence in the $y$ direction, and $\omega$ is the complex angular frequency of the wave. In addition to the free surface boundary conditions at $y=0$,
the boundary conditions of velocity continuity and stress balance are imposed at the water-oil interface at $y=-h_w$ (see SI Appendix, section 1). The real part of $\omega$ corresponds to the oscillation frequency $\omega_0$, and the imaginary part corresponds to the damping rate $T_\lambda^{-1}$. 
$\omega$ is determined based on specified boundary conditions. $T_\lambda^{-1}$ is then solved (see SI Appendix, section 2) when the oil thickness and wavelength are given, which will be characterized in the next section.

\section*{Capillary wave modeling}

\begin{figure}
    \centering
    \includegraphics[width=0.85\linewidth]{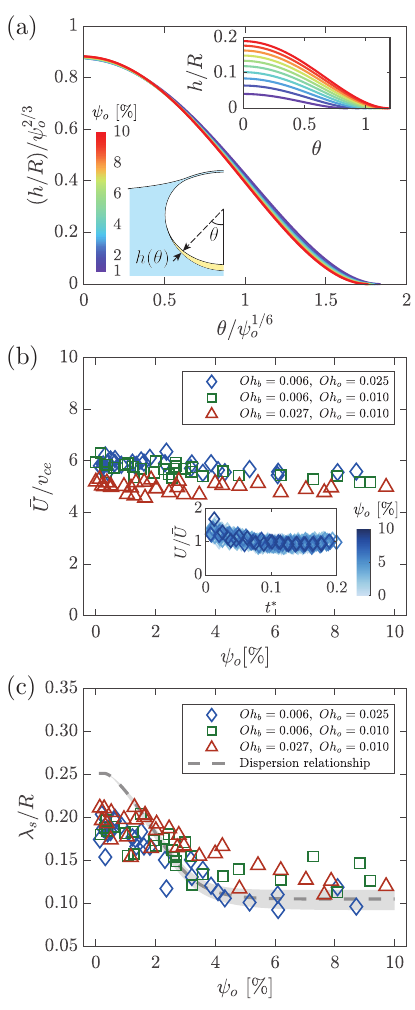}
    \caption{(a) Collapse of rescaled initial oil thickness, $h/(R\psi_o^{2/3})$, versus rescaled polar angle, $\theta/\psi_o^{1/6}$. The inset shows the initial dimensionless oil thickness $h/R$ as a function of the polar angle $\theta$.  (b)  The dimensionless average velocity $\bar{U}/v_{ce}$ as a function of oil fraction $\psi_o$. The inset shows the evolution of normalized wave velocity $U/\Bar{U}$ with dimensionless time $t^* = t/(R/v_{ce})$. (c) Dimensionless wavelength of the secondary precursor wave $\lambda_s/R$ at $\theta = \pi/4$ as a function of oil fraction $\psi_o$. The wavelengths calculated from Eq. (\ref{eq:Ubt}) and $h(\theta)$ using the average experimental parameters of liquids are shown as the grey dashed line, and the shaded area represents the range of wavelengths calculated across the full range of experimental parameters (see SI Appendix, section 5).}
    \label{fig:wavemodel}
\end{figure}

\textbf{Oil layer thickness profile.}
To determine the oil layer thickness $h_w$ for the wave damping model,
we calculate its initial distribution at the cavity surface of compound bubbles.   
The oil-coated bubble shape calculated using the Young-Laplace equations \cite{lhuissier2012bursting,yang2023enhanced} shows good agreement with experimental results as shown by the red dashed curves in Fig. \ref{fig:cavity}(a). 
The inset of Fig. \ref{fig:wavemodel}(a) illustrates the dimensionless oil layer thickness $h(\theta)/R$ at the static bubble cavity surface for different oil fractions. 
We note that the oil layer thickness obeys the scaling relations of 
$h(\theta)/R\sim\psi_o^{2/3}$ and $\theta\sim\psi_o^{1/6}$. 

Next, we obtain $h_w$ from the initial shape of the oil layer and the wave velocity. With the predicted oil layer distribution, we determine the local oil thickness $h_w$ encountered by the capillary wave based on the propagation speed of the waves. We measure the average wave velocity $\bar{U}$ for selected cases by tracking the position of the dominant wave trough when $\theta>\pi/6$, as shown in Fig. \ref{fig:wavemodel}(b). For different oil Ohnesorge numbers $Oh_o=\mu_o/\sqrt{\rho_o\gamma_{oa}R}$ and bulk Ohnesorge numbers $Oh_b=\mu_w/\sqrt{\rho_w(\gamma_{oa}+\gamma_{ow})R}$, the dimensionless average wave velocity $\Bar{U}/v_{ce}$ remains a nearly constant value of $5.7\pm0.7$ regardless of the oil fraction $\psi_o$. Here, the effective capillary velocity is defined as $v_{ce}=\sqrt{(\gamma_{oa}+\gamma_{ow})/(\rho R)}$. Such a constant wave velocity enables us to convert the spatial distribution of $h(\theta)$ into the time dependency of $h_w(t)$ for our further calculation of a varying damping rate.

\noindent \textbf{Wavelength of secondary wave.} To obtain $\lambda_s$ for solving $T_\lambda^{-1}$, the wave dispersion relationship must be considered since capillary waves traveling at a constant velocity $U$ should adopt fixed wavelengths determined by the dispersion relationship $\omega_0(k)$ \cite{lighthill2001waves}. Considering $\lambda_s \gg l_v$, the dispersion relationship of the capillary waves can be described solely by the irrotational motion of the wave \cite{jeng1998viscosity,denner2016frequency}. The irrotational analysis of the capillary wave model yields two analytical solutions for wave dispersion, corresponding to the sinuous (barotropic) and varicose (baroclinic) modes documented in the literature \cite{rajan2020three,stewart2015} (see SI Appendix, section 3). 
Because the wave motion at the compound interface is initiated in phase (Figs. \ref{fig:images}(a) and \ref{fig:cavity}(a)), we focus on the sinuous mode which obeys the dispersion law of
\begin{equation} \label{eq:Ubt}
    \frac{U_{bt}}{\sqrt{\gamma_{oa}k/\rho_o}}=\frac{1}{2}\sqrt{\Upsilon_1+\sqrt{\Upsilon_1^2-4\Upsilon_0}},
\end{equation}
where
\begin{equation*}
\left\{
\begin{array}{l}
    \Upsilon_1 = \frac{\me^{-2kh_w}(1-\mathrm{P}^{-1})+1+\mathrm{P}^{-1}+\mathrm{Y}^{-1}(1+\me^{-2kh_w})}{\me^{-2kh_w}(\mathrm{P}^{-1}-1)+1+\mathrm{P}^{-1}},\\
    \Upsilon_0 = \frac{\mathrm{Y}^{-1}(1-\me^{-2kh_w})}{\me^{-2kh_w}(\mathrm{P}^{-1}-1)+1+\mathrm{P}^{-1}}.
\end{array}
\right.
\end{equation*}
Here, $U_{bt}$ is the phase speed of the sinuous mode, $\mathrm{P}=\rho_o/\rho_w$, and $\mathrm{Y}=\gamma_{oa}/\gamma_{ow}$.
Substituting $U_{bt}=\bar{U}$ into Eq. (\ref{eq:Ubt}), the wavelength $\lambda_s = 2\pi /k$ is uniquely determined for each $h_w$.

The wavelength obtained from Eq. (\ref{eq:Ubt}) shows how the oil layer thickness influences the precursor wave.
Two limiting cases of $h_w\rightarrow 0$ and $h_w\rightarrow\infty$ are recovered for the secondary wavelength for oil-coated bubble bursting using 
$\lambda_s\approx 2\pi\gamma/(\rho {U}^2)$ from the wave dispersion relation at a bare surface. When $h_w\rightarrow 0$, the oil-air and oil-water interfaces oscillate synchronously at the water surface with $\rho = \rho_w$ and $\gamma = \gamma_{oa}+\gamma_{ow}$. The wavelength from Eq. (\ref{eq:Ubt}) gives $\omega_0 = \sqrt{(\gamma_{oa}+\gamma_{ow})k^3/\rho_w}(1+O(kh_w))$ and $\lambda_s \approx 0.25R$, consistent with previous theoretical results \cite{rajan2020three} (see SI Appendix, section 3) and the wavelengths of the secondary wave observed for bare bubble bursting \cite{yang2023daughter}. 
In contrast, when $h_w\gg \lambda_s$, the capillary waves on the two interfaces are spatially separated as shown in Fig.~\ref{fig:images}(b). Thus, the capillary waves at the top interface effectively travel in a deep layer of oil with $\rho = \rho_o$ and $\gamma=\gamma_{oa}$, resulting in $\lambda_s \approx 0.1R$.
In Fig. \ref{fig:wavemodel}(c), the wavelengths calculated using Eq.~(\ref{eq:Ubt}) with experimental parameters (see SI Appendix, section 5) are compared against experimental measurements at $\theta = \pi/4$. The theoretical predictions capture the experimental trend of wavelength variation with the oil fraction, which are also consistent with the above two limiting cases. The modified dispersion relationship enables us to estimate $\lambda_s$ for any given $h_w$, which allows us to further solve for the instantaneous damping rates of the capillary waves.

\noindent \textbf{$Oh_r$ for cumulative damping of precursor waves.}
The evolution of the instantaneous damping rate determines the cumulative damping of the capillary wave along the cavity surface.  
During the propagation of capillary waves, their energy $E$ attenuates due to viscous dissipation.
Since the dissipation of the total energy $E$ of the capillary wave follows 
\cite{lamb1924hydrodynamics,ermakov2022resonance}
\begin{equation}
    \frac{\mathrm{d}E}{\mathrm{d}t} = 2T_\lambda^{-1}E,
\end{equation}
the logarithmic damping on the wave energy from the initial energy $E_0$ to the final energy $E_f$, representing a cumulative damping effect, can be expressed as \cite{longuet1997viscous,krishnan2017scaling}
\begin{equation}
    \log\left(\frac{E_f}{E_0}\right) = 2\int_0^{t_{bc}} T_\lambda^{-1}\mathrm{d}t.
\end{equation}
For bare bubbles where $T_\lambda^{-1}$ is a constant, the logarithmic damping simplifies to $\log(E_f/E_0) = 2T_\lambda^{-1}t_{bc}\propto Oh$, which connects $Oh$ to the cumulative damping effect. 
We thus define the revised Ohnesorge number for oil-coated bubbles where $T_\lambda^{-1}$ varies with time as
\begin{equation}
    Oh_r = 2K\int_0^{t_{bc}} T_{\lambda}^{-1}\mathrm{d}t,
\end{equation}
where the coefficient $K\approx0.0013$ (see SI Appendix, section 4). 
Notably, the revised Ohnesorge number accounts for the viscous damping introduced by a non-uniform oil layer in oil-coated bubble bursting.

\section*{Predicting jet size with revised Ohnesorge number}

\begin{figure}[!h]
    \centering
    \includegraphics[width=0.85\linewidth]{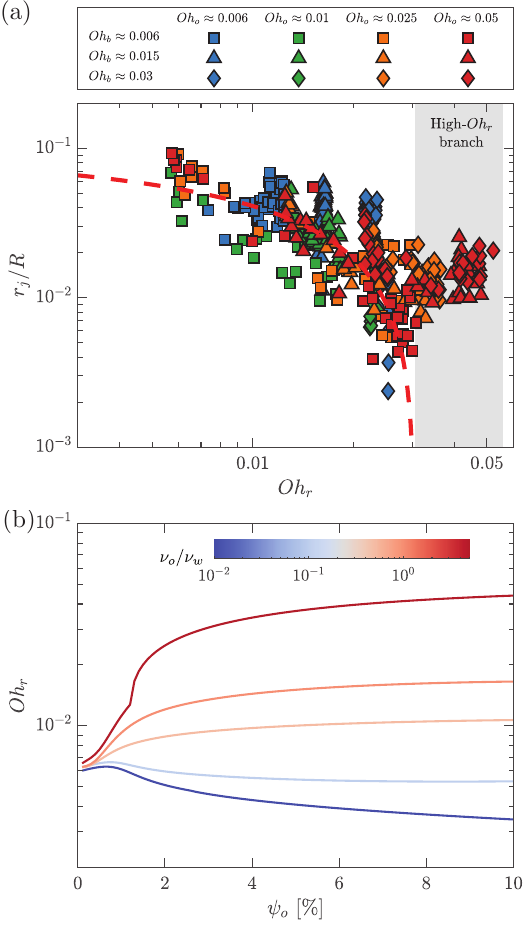}
    \caption{(a) Dimensionless jet radius $r_j/R$ as a function of revised Ohnesorge number $Oh_r$. The red dashed line represents the modified $Oh_r - r_j/R$ scaling relation of   ${r_j}/{R} = K_j(1-\sqrt{Oh_{r}/0.0305})$, where $K_j= 0.096 $ is a parameter from least square fitting. The shaded area indicates the high $Oh_r$ regime where the viscous effect suppresses jet growth and a different scaling law may hold. (b) Revised Ohnesorge number $Oh_r$ as a function of $\psi_o$ at $Oh_b = 0.006$ and different $\nu_o/\nu_w$ (see SI Appendix, sections 4 and 5 for details).  }
    \label{fig:collapse}
\end{figure}

Figure \ref{fig:collapse}(a) shows the dimensionless jet radius $r_j/R$ as a function of the proposed $Oh_r$. The jet radius decreases as the revised Ohnesorge number increases, showing that a smaller jet radius is expected with stronger wave dissipation represented by higher $Oh_r$. Considering the scaling relation of $r_j/R$ versus $Oh$ for bare bubble bursting \cite{blanco2021jets}, we propose a modified scaling relation by substituting $Oh$ with $Oh_r$ for a compound multiphase bubble as follows:
\begin{equation}\label{eq:rjovr}
    {r_j}/{R} \propto 1-\sqrt{Oh_{r}/0.0305}.
\end{equation}
For our experiments covering a wide range of bulk Ohnesorge numbers $Oh_b$ from 0.006 to 0.03 and oil Ohnesorge numbers $Oh_o$ from 0.006 to 0.05, the proposed scaling law reasonably well captures the experimentally observed jet radii spanning two orders of magnitude. The consistency between the experimental results and the scaling law indicates that $Oh_r$ effectively characterizes the influence of the compound layer on the jet radius.  
In addition, we note the scaling law only holds for Ohnesorge numbers below a critical $Oh_c=0.0305$ \cite{ganan2021physics,blanco2021jets}, beyond which the viscous effect limits the minimum jet radius during its growth as found in bare bubble bursting\cite{brasz2018minimum,blanco2020sea}. 

Figure \ref{fig:collapse} (b) shows the revised Ohnesorge number as a function of different oil fractions with the same $Oh_b = 0.006$ and varying kinematic viscosity ratios. Notably, when the oil viscosity is greater than or comparable to the water viscosity ($\nu_o/\nu_w = 5, 1$ or $0.5$), $Oh_r$ increases with increasing oil fraction, indicating stronger damping as $\psi_o$ rises. 
In contrast, when the oil viscosity is significantly lower than the water viscosity ($\nu_o/\nu_w = 0.1$ or $0.01$), our model predicts that $Oh_r$ decreases with $\psi_o$. To understand how the $Oh_r-\psi_o$ trend depends on the kinematic viscosity ratio, we express the Ohnesorge number as 
\begin{equation}
Oh \propto T_\lambda^{-1}t_{bc} \propto \frac{\nu t_c}{\lambda_s^2},
\end{equation}
since $T_\lambda^{-1}\propto \nu\lambda_s^{-2}$ and $t_{bc}\propto t_c$. Considering $\delta = \sqrt{\nu t_c}$ represents the characteristic boundary layer thickness \cite{batchelor2000introduction,tang2019spreading,gordillo2019capillary}, $Oh$ can indicate the ratio between the boundary layer length scale and the secondary wavelength. 
For oil-coated bubble bursting, the oil layer thus influences $Oh_r$ by altering both the boundary layer thickness and the wavelength of the precursor waves. Increasing $\psi_o$ decreases $\lambda_s$ from $\approx$$0.25R$ to $0.1R$ as shown in Fig.~\ref{fig:wavemodel}(c). Meanwhile, the characteristic $\delta$ depends on the kinematic viscosities of both liquids at the compound interface, with oil viscosity contributing more as $\psi_o$ increases. When $\nu_o > \nu_w$, $\delta$ increases with $\psi_o$ since a thicker oil layer with a higher oil kinematic viscosity allows a larger depth of penetration for viscous diffusion. Therefore, the increase in $\delta$ and the decrease in $\lambda_s$ collectively contribute to an increase in $Oh_r$ as $\psi_o$ increases. However, when $\nu_o < \nu_w$, $\delta$ decreases with increasing $\psi_o$ given smaller $\nu_o$. When the decrease in $\delta$ outweighs the decrease of $\lambda_s$ and thus decreases $Oh_r$, the precursor waves essentially experience less damping by the time they reach the cavity bottom. The undamped perturbations from the precursor waves disrupt the focusing of the dominant waves and cause the jet size to increase. While our revised Ohnesorge number describes the jet radius well, the jet velocity will require further discussion. For the same $Oh_r$, we observe that the jet velocity decreases with $Oh_o$. The phenomenon may originate from the additional viscous stresses due to the multiphase flow during jet growth, which is beyond the scope of the current work.

We believe that the introduction of $Oh_r$ provides a fundamental framework to characterize how a structurally compound liquid-bubble interface influences the microscale fluid mechanics associated with bubble dynamics. Specifically, bubbles are often contaminated by surface-active substances such as surfactants, proteins, or particulates, which are prevalent in various natural and industrial settings \cite{wurl2011formation,walls2014moving,walls2017,poulain2018b,neel2021collective}. Such contaminated interfaces may exhibit diverse stress responses to deformation, including Marangoni stress, surface viscosity and surface elasticity. Our methodology based on capillary wave theory may be broadened to include a wide variety of these systems, offering a robust approach to characterizing the resulting alterations in jetting dynamics. 

\section*{Conclusions}

In this study, we have investigated the influence of an immiscible compound coating on the size of bubble bursting jets. We identified that the damping of capillary waves during their propagation plays a critical role in determining jet characteristics. As the oil fraction increases, the impact of the oil-water interface on precursor waves gradually diminishes as they propagate to the cavity bottom, while the oil viscosity becomes increasingly significant, leading to a reduction in the characteristic wavelength. To account for these effects, we developed a linearized wave model to calculate the damping rate of capillary waves. A revised Ohnesorge number that incorporates the influence of the compound interface on the capillary wave damping was proposed, which 
effectively collapses the jet size data across a wide range of oil coating fractions and viscosity ratios. Our scaling relation for the jet size offers model constraints and engineering insights for the ejected contaminant-laden drop size, which can be applied to evaluate bubble-mediated airborne flux of pollutants from multiple sources including ocean or wastewater treatment. In addition, our current study focuses on bubbles coated with a Newtonian fluid layer, establishing a benchmark framework for future work that investigates compound bubbles coated with rheologically complex fluids. Given the prevalence of complex biocontaminants, such as extracellular polymeric substances produced through microbial secretion, metabolism, and lysis in natural and engineered systems \cite{more2014extracellular, poulain2018}, understanding the bursting dynamics of bubbles coated with a non-Newtonian compound layer would be of significant fundamental interest and hold broader implications for bubble-mediated marine chemistry and biology as well as environmental science.

\bigskip
\textbf{Acknowledgements.} Z.Y. and J.F. acknowledge partial support by the National Science Foundation (NSF) under grant  No. CBET 2426809 and the research support award RB24105 from Campus Research Board at University of Illinois at Urbana-Champaign. We also thank Bingqiang Ji for fruitful discussions about the experiments.

\section{Methods}
\subsection{Materials} 

Deionized water (electrical resistivity = 18.2 M$\rm \Omega\cdot$cm at $25^\circ$ C) is produced by a laboratory water purification system (Smart2Pure 3 UV/UF, Thermo Fisher Scientific). Silicon oils including octamethyltrisiloxane ($\nu$ = 1 cSt), dodecamethylpentasiloxane ($\nu$ = 2 cSt) and others ($\nu$ = 5, 10 cSt) were purchased from Sigma-Aldrich and used as received. Glycerin was purchased from Fisher Chemical. The surface tensions were measured using the pendant drop method \cite{song1996determination}, including the surface tension between the oil and air ($\gamma_{oa}$), the glycerin solution and air ($\gamma_{wa}$), and the oil and glycerin solution ($\gamma_{ow}$) (see SI Appendix, section 5).

\subsection{Experimental setup} 

The oil-coated bubbles were generated 
by a coaxial orifice system plugged in an acrylic container, a setup described in detail in our previous work \cite{ji2021oil,yang2023enhanced}. The equilibrium radius of the released compound bubble (gas+oil) in our experiments was determined to be $R \approx 2$ mm. After the bubbles rested at the water surface, we used two high-speed cameras (FASTCAM Mini AX200, Photron) to synchronously record the top and bottom side views of the bubble cavity collapse and jet formation, at a frame rate of 6400-20000 frames per second and a magnification of 1-4. Two LED panels (Phlox 100$\times$100 HSC; Phlox 200$\times$200 HSC) were used to illuminate the bubble bursting for the two cameras separately. The obtained images were post-processed with Fiji/ImageJ and MATLAB.

\bibliographystyle{prsty_withtitle}
\bibliography{sn-bibliography}

\end{document}